\documentclass[11pt,a4paper,twocolumn]{article}
\usepackage[utf8x]{inputenc}
\usepackage[T1]{fontenc}
\usepackage{mathptmx} 

\usepackage{balance} 
\usepackage{algorithm}
\usepackage{algpseudocode}

\usepackage{amsthm}
\usepackage{amsmath}
\usepackage{amsfonts}
\usepackage{graphicx}
\usepackage{comment}
\usepackage{url}
\usepackage{authblk}

\theoremstyle{definition}
\newtheorem{definition}{Definition}[section]

\newtheorem{theorem}{Theorem}[section]

\usepackage{xcolor} 

\begin{document}

\title{ATL$^*$AS: An Automata-Theoretic Approach and Tool\\ for the Verification of Strategic Abilities in Multi-Agent Systems}


\author{Sofia Garcia de Blas Garcia-Alcalde \and Francesco Belardinelli}
\affil{Department of Computing, Imperial College London}








         







\maketitle


\begin{abstract}
We present two novel symbolic algorithms for model checking
the Alternating-time Temporal Logic ATL$^*$, over both the infinite-trace and the finite-trace semantics.
In particular, for infinite traces we design a novel symbolic reduction to
parity games. 
We implement both methods in the ATL$^*$AS model checker and evaluate it using synthetic
benchmarks as well as a cybersecurity scenario. Our results demonstrate that the symbolic approach
significantly outperforms the explicit-state representation and we
find that our parity-game-based algorithm offers a more scalable and
efficient solution for infinite-trace verification, outperforming previously available tools.
Our results also confirm that finite-trace
model checking yields substantial performance benefits over
infinite-trace verification. As such, we provide a comprehensive toolset for
verifying multi-agent systems against specifications in ATL$^*$.
\end{abstract}


\section{Introduction}

Formal verification of multi-agent systems (MAS) is essential for ensuring
correctness in strategic, high-stakes settings \cite{dorri_2018_multiagent}.
Alternating-time Temporal Logic was originally proposed as a specification language to express strategic abilities of agents in MAS \cite{alur_2002_alternatingtime}.
Since then, it has been extended in multiple directions to account for specific features of MAS, including imperfect information and knowledge \cite{JamrogaH04}, stochastic transitions \cite{ChenL07a}, quantification on strategies \cite{MogaveroMPV14,ChatterjeeHP10}, among others.

A key distinction in this domain is about modelling system executions
as either infinite or finite traces. The tradition of reactive systems
has usually considered infinite traces \cite{Pnueli86}, suitable to model the system's
readiness to accept inputs from the user.
Nonetheless, 
%
many scenarios -- including games and learning algorithms --
naturally lend themselves to finite-trace modelling, where executions
terminate after a finite number of steps.
This has led to a vast literature translating results available for the infinite-trace setting to the finite-trace case \cite{degiacomo_2013_linear,degiacomo_2015_synthesis,belardinelli_2018_alternatingtime}. 

Besides modelling purposes, finite traces also offer practical advantages regarding the corresponding verification procedures. 
Specifically, model
checking over finite traces might avoid the complexities of \(\omega\)-automata and instead use finite automata, yielding simpler algorithms \cite{belardinelli_2018_alternatingtime}.

\paragraph{Contributions.}
We 
advance the state of the art in the formal verification of
multi-agent systems by developing, implementing and evaluating effcient, symbolic model checking
algorithms for both finite- and infinite-trace variants of ATL$^*$. The
core contributions of this work are:
\begin{enumerate}

\item \textit{A symbolic algorithm  for
  $\text{ATL}_f^*$ model checking and its implementation},
  built on the theoretical
  foundations and complexity result presented in
  \cite{belardinelli_2018_alternatingtime}. Our symbolic algorithm
  includes several design choices that improve
  substantially over \cite{belardinelli_2018_alternatingtime}
  and enable scalable
  implementation. This is accompanied
  by the full implementation of both the explicit and symbolic model
  checking algorithms, thus achieving the first tool for
  verifying $\text{ATL}_f^*$.

\item
  \textit{A symbolic reduction of ATL\(^*\) model checking to parity games}, resulting in the first available implementation for full ATL\(^*\) over infinite traces.

\item\textit{Evaluation on a strategic cybersecurity scenario:} In Sec.~\ref{sec:modelling}, a concrete attacker-defenders MAS is modelled to demonstrate the applicability of the symbolic $\text{ATL}_f^*$ model checker to realistic verification tasks.

\item \textit{Comparative analysis of finite vs. infinite-trace
  verification:}
  We investigate empirically the practical
  performance trade-offs between finite- and infinite-trace ATL\(^*\) model
  checking, as well as between different algorithms in both
  domains. The results of these evaluations are presented in
  Sec.~\ref{sec:evaluation}.
\end{enumerate}

These algorithms are integrated into ATL\(^*\)AS, a C++ model checking tool that supports both finite and
infinite-trace verification of ATL\(^*\). Users can provide a system
specification and select among the provided algorithms to verify temporal and strategic properties efficiently. The code of ATL\(^*\)AS is provided in the supplementary material to this paper.

\paragraph{Related Work.}
The computational complexity of model checking temporal logics varies significantly depending on the language considered and its expressive power. Table~\ref{tab:model_checking_complexity} summarises the complexity results for model checking some key multi-agent logics, under the assumption of perfect recall strategies.
Existing work on ATL\(_f^*\) model checking is centered around the explicit-state algorithm introduced in \cite{belardinelli_2018_alternatingtime}, which provides the first decidability and complexity results for model checking ATL\(^*_f\). However, this approach suffers from limited scalability due to full state-space enumeration. ATL\(^*\) over infinite traces lacks dedicated tools, but can be indirectly verified via the SL[1G] fragment of strategy logic, supported by the symbolic MCMAS-SL[1G] checker \cite{cermk_2014_sl1g}, which uses generalised parity games. Parity-game-based model checking has also been developed for logics such as the modal \(\mu\)-calculus \cite{emerson_2001_on} and game logic \cite{landsaat_2022_a}. The reduction of ATL\(^*\) model checking to parity games was established in \cite{alfaro_2002_from}, showing that control problems over Kripke structures with \(\omega\)-regular objectives can be encoded as parity games. However, no concrete algorithm for concurrent game structures was provided, nor were the symbolic aspects required for efficient practical implementation addressed.
\begin{table}[H]
\centering
\caption{Model Checking Complexity of Relevant Logics}
\label{tab:model_checking_complexity}
\begin{tabular}{|p{.8cm}|p{3.3cm}|p{3cm}|}
\hline
\textbf{Logic} & \textbf{Complexity} & \textbf{Implementation} \\
\hline
ATL & PTIME-complete \cite{alur_2002_alternatingtime} & MCMAS \cite{lomuscio_2015_mcmas} \\
\hline
ATL\(^*\) & 2EXPTIME-complete \cite{alur_2002_alternatingtime} & {\bf ATL$^*$AS (here)} \\
\hline
$\text{ATL}_f^*$ & 2EXPTIME-complete \cite{belardinelli_2018_alternatingtime} & {\bf ATL$^*$AS (here)} \\
\hline
SL & Non-elementary \cite{mogavero_2014_reasoning} & MCMAS-SLK \cite{cermk_2014_mcmasslk} \\
\hline
SL[1G] & 2EXPTIME-complete \cite{mogavero_2014_reasoning} & MCMAS-SL[1G] \cite{cermk_2014_sl1g} \\
\hline
\end{tabular}
\end{table}
\section{Background}
In this section we introduce the background on Alternating-time
Temporal Logic, including its syntax and semantics (Sec.~\ref{sec:atl}) and the basics on automata theory to solve the corresponding model checking problem (Sec.~\ref{sec:automata}).
In what follows, give a (possibly infinite) sequence $\tau = t_0, \ldots, t_n, \ldots$ of elements from set $T$, we denote its length as $|\tau| \in \mathbb{N} \cup \{\infty \}$, the $i$-th element as $\tau_i = t_i$, and the subsequence $t_i, \ldots, t_{n+i}, \ldots$ starting at position $i$ as  $\tau_{\geq i}$.

\subsection{Alternating-time Temporal Logic}
\label{sec:atl}
\paragraph{Syntax.}
We state the syntax for the language ATL\(^*\) as introduced in \cite{alur_2002_alternatingtime}, which is also the syntax for its variant  ATL\(^*_f\)
interpreted over finite traces.
Let $AP$ be the set of {\em atomic propositions} (or atoms) and $Ag$ the set of {\em agents}.  Formulas in ATL\(^*\) are the
state  formulas $\varphi$ built using path formulas $\psi$ with Boolean operators and strategic quantifiers:
\begin{eqnarray*}
    \varphi & ::= & 
    p \mid \neg \varphi \mid \varphi_1 \land \varphi_2 \mid \langle\!\langle A\rangle\!\rangle \psi\\
        \psi & ::= & \varphi \mid \neg \psi \mid \psi \land \psi  \mid \mathsf{X} \psi \mid \psi \mathsf{U} \psi
\end{eqnarray*}
where $p \in AP$ and $A \subseteq Ag$.


Operator {\em next} $\mathsf{X}$ and {\em until} $\mathsf{U}$ are the standard modalities of Linear-time Temporal Logic (LTL) \cite{pnueli_1977_the}.
Additional temporal operators are also derivable from $\mathsf{X}$ and $\mathsf{U}$, including finally $\mathsf{F} \varphi \equiv \texttt{true} \mathsf{U} \varphi$ and globally $\mathsf{G} \varphi \equiv \neg \mathsf{F} \neg\varphi$.
For simplicity, we include these derived operators in the syntax.

\paragraph{Semantics.} We present the semantics of ATL\(^*\) by means of Concurrent Game Structures.
\begin{definition}[CGS]
    \label{CGS}
    A {\em Concurrent Game Structure} is a tuple $G =\langle Ag, AP,S, s_0, {Act}_{a \in Ag},\delta,\lambda \rangle $ where:
    \begin{itemize}
        \item $Ag$ is the finite, non-empty set of {\em agents};
        \item $AP$ is the finite, non-empty set of {\em atoms};
        \item $S$ is the finite, non-empty set of {\em states}, with {\em initial state} $s_0 \in S$;
        \item for each agent $a \in Ag$, $Act_a$ is a finite, non-empty set of {\em actions};
        \item $\delta : S\times Jact\rightarrow S$ is the {\em transition function}, where $Jact = \prod_{a\ \in\ Ag}{Act}_a$ is the set of joint actions;
        \item $\lambda:S\rightarrow2^{AP}$ is the {\em labelling function} that assigns to each state $s$ the set $\lambda(s)$ of atoms that are true in $s$.
    \end{itemize}
\end{definition} 

To interpret ATL\(^*\) on finite traces, we extend Def.~\ref{CGS} with a set $F \subseteq S$ of final states to identify terminating executions, as originally proposed in \cite{belardinelli_2018_alternatingtime}.

Further, we define the following concepts related to CGSs.
%
A \textit{path} is a sequence $\pi= s_0, s_1, \ldots, s_n, \ldots$ of states, either infinite, or finite terminating in a final state in $F$, where 
for every $i \le |\pi|$, there exists a joint action $J \in Jact$ such that $\delta\left(\pi_i,J\right)=\pi_{i+1}$.
In the finite-trace setting, we additionally define a \textit{history} as a finite path not necessarily ending in a final state. 
The sets of all histories and paths are denoted as $Hist$ and $Path$ respectively.

A \textit{strategy} for an agent $a\in Ag$ is a function $\sigma_a : Hist \rightarrow{Act}_a$, mapping histories to actions. 
Notice that here we consider \textit{perfect recall} strategies, as action selection depends on the entire history, not just on the current state. The set of all (perfect recall) strategies is denoted as $\sum_{R}{(G)}$.
A \textit{joint strategy} $\sigma_A : A \rightarrow \sum_{R}{(G)}$ is a function associating  a strategy to each agent in coalition $A \subseteq Ag$.

Given a state $s \in S$, the \textit{finite} (resp.~\textit{infinite}) outcome of a joint strategy $\sigma_A$ from $s$, denoted $\text{out}_f\left(s,\sigma_A\right)$ (resp.~$\text{out}_\infty(s, \sigma_A)$), is the set of finite (resp.~infinite) paths  $\pi$ consistent with $s$ and $\sigma_A$, that is,
(i) $\pi_0=s$;
and (ii) for every $i< |\pi|$, there exists a joint action $J \in Jact$ such that $\pi_{i+1} = \delta\left(\pi_i,J\right)$ and $J(a)=\sigma_A(a)(\pi_{\le i})$ for each agent $a \in A$.
We now define a parametrised semantics for \( x \in \{\infty, f\} \) to denote the infinite and finite trace interpretation, respectively.

{\small
\begin{tabbing}
$(G, s) \models_x p$  \  \ \ \ \ \ \ \ \ \ \ \ \ \= iff \ \= $p \in \lambda(s)$ \\
$(G, s) \models_x \neg \varphi$ \> iff \> $(G, s) \not\models_x \varphi$\\
$(G, s) \models_x \varphi \land \varphi'$ \> iff \> $(G, s) \models_x \varphi$  and $(G, s) \models_x \varphi'$ \\
$(G, s) \models_x \langle\!\langle A \rangle\!\rangle \psi$ \> iff \> $\exists \sigma_A \in \Sigma_R(G)$ s.t. \\
\> \> $\forall \pi \in \text{out}_x(s, \sigma_A),\ (G, \pi) \models_x \psi$ \\
$(G, \pi) \models_x \varphi$ \> iff \> $(G, \pi_0 \models_x \varphi$\\
$(G, \pi) \models_x \neg \psi$ \> iff \> $(G, \pi) \not\models_x \psi$ \\
$(G, \pi) \models_x \psi_1 \land \psi_2$  \> iff \> $(G, \pi) \models_x \psi_1$ and  $(G, \pi) \models_x \psi_2$ \\
$(G, \pi) \models_x \mathsf{X} \psi$  \> iff \> $|\pi| > 1$ and $(G, \pi_{\geq 1}) \models_x \psi$ \\
$(G, \pi) \models_x \psi_1\ \mathsf{U}\ \psi_2$  \> iff \> $\exists j < |\pi|$ s.t. $(G, \pi_{\geq j}) \models_x \psi_2$ and \\
\> \> for all $0 \leq k < j$, $(G, \pi_{\geq k}) \models_x \psi_1$
\end{tabbing}
}

Notice that on infinite traces, the condition for $\mathsf{X} \psi$ whereby $|\pi| >1$ is always satisfied, as $|\pi| = \infty$.

We write \( G \models_x \varphi \) as shorthand for \( (G, s_0) \models_x \varphi \), where \( s_0 \) is the initial state of \( G \).



\paragraph{Model Checking.} In the rest of this paper, we are interested in the following model checking problem.
\begin{definition}[Model Checking Problem]
    Given a CGS $G$ and an $ATL^*$ formula $\phi$, determine whether $G \models \phi$.
\end{definition}

Complexity results for model checking ATL$^*$, relevant fragments and extensions are reported in Table~\ref{tab:model_checking_complexity}, for the case of perfect recall strategies.

\subsection{Automata}
\label{sec:automata}
Our approach relies on automata-theoretic techniques for model checking temporal logics. Our algorithms use deterministic finite automata (DFAs) to represent LTL\(_f\) formulas in the finite trace interpretation, and deterministic parity automata (DPAs) to represent LTL formulas when interpreted over infinite traces. We define both these automata hereafter.
\begin{definition}[DFA]
    \label{Deterministic Finite Automaton}
    A {\em deterministic finite automaton} is a tuple $\mathcal{A} =\langle Q, q_0, \Sigma,\delta,F\rangle$ where
  (i) $Q$ is a finite set of {\em states}, with {\em initial state} $q_0$;
  (ii) $\Sigma$ is the {\em alphabet};
  (iii) $\delta:Q\times\Sigma\rightarrow Q$ is the {\em transition function};
   (iv) $F \subseteq Q$ is a set of {\em final states}. 
\end{definition}
A {\em run} $q$ on a DFA for a finite word $w \in \Sigma^*$ is a finite sequence $q_0q_1, \ldots, q_n$ of states such that for $i < |w|$, $q_{i+1} = \delta(q_i, w_i)$. A run is  {\em accepting} if it ends in a final state, i.e., $q_n \in F$.

\begin{definition}[DPA]
    \label{Deterministic Parity Automaton}
    A {\em deterministic parity automaton} is a tuple $A =\langle Q, q_0, \Sigma,\delta, F\rangle$, where
    $Q, q_0, \Sigma, \delta$ are defined as for DFAs, and $F$ is now a parity acceptance condition $\{P_1,P_2, \dots, P_n\}$, where each $P_i \subseteq S$. Moreover, an infinite run $\sigma = q_0q_1\dots$ is {\em accepting} if
$\min_i \left(P_i \cap \text{Inf}(\sigma) \neq \emptyset \right)$ is odd, where $\text{Inf}(\sigma)$ is the set of states occurring infinitely often in $\sigma$.
\end{definition}

Given an automaton $\mathcal{A}$, the {\em language} $\mathcal{L}(\mathcal{A})$ is the set of words that are accepted by some run in $\mathcal{A}$.
Deterministic parity automata induce a natural class of two-player games, known as parity games, where acceptance conditions are rephrased in terms of winning strategies over infinite plays.
\begin{definition}[Parity Game]
A (min‐){\em parity game} is a tuple $\mathcal{G} = (V,\,V_0,\,V_1,\,E,\,\Omega)$,
where \((V,E)\) is a finite directed graph, \(V=V_0\cup V_1\) partitions positions between player ~0 (Even) and player ~1 (Odd), and \(\Omega\colon V\to\mathbb{N}\) assigns each vertex a \emph{priority}.  A play is an infinite path \(v_0v_1v_2\cdots\); player~0 wins if the minimum priority occurring infinitely often is even, otherwise player~1 wins.
\end{definition}


\section{Symbolic Algorithm for Finite Traces}
\label{sec:finite-algs}
The model checking procedure for ATL\(^*_f\) follows a 
recursive labelling approach, where subformulas are evaluated bottom-up and annotated over the state space, as described in \cite{belardinelli_2018_alternatingtime}. The key computational step occurs when evaluating a strategy formula of the form $\langle\!\langle A\rangle\!\rangle \psi$. Here, the path formula $\psi$ is first rewritten into an equivalent $\text{LTL}_f$ formula by replacing all maximal state subformulas with fresh atoms. The core task then reduces to computing the set of states from which coalition $A$ can enforce $\psi$. This is solved by invoking the \texttt{GameSolving} procedure, whose high-level structure is shown in Algorithm~\ref{alg:game-solving-finite}.

The algorithm first translates the LTL\(_f\) formula into an equivalent DFA (line 1) \cite{degiacomo_2013_linear}. The product between the CGS $G$ and the DFA is then constructed as per Def.~\ref{def:product-graph} (line 2), yielding a structure that simultaneously tracks the evolution of the MAS and the progress towards satisfying $\psi$. To initialise the fixpoint computation, an initial safety set is derived in line 3, it includes all product states that are either not final in the CGS or are accepting in the DFA. Intuitively, this set excludes precisely those states in which the system has terminated and the property is not yet satisfied. Finally, the greatest fixpoint iteration (line 4) computes the full set of CGS states from which the coalition can enfoce $\psi$ through some joint strategy. We present novel symbolic algorithms for the two central steps in this procedure: product construction and 
fixpoint solution.
\begin{algorithm}
\caption{\texttt{GameSolving}}
\label{alg:game-solving-finite}
\begin{algorithmic}[1]
\Statex \textbf{Input:} CGS $G$, Strategic formula $\langle\!\langle A\rangle\!\rangle \psi$
\Statex \textbf{Output:} $\{q \in S \; | \; (G, q) \models \langle\!\langle A \rangle\!\rangle \psi \}$
\Statex
\State Convert LTL$_f$ formula $\psi$ into DFA $\mathcal{A}_\psi$
\State Construct product graph of $G$ and $\mathcal{A}_\psi$
\State Derive safety set $Safe = \{(q, s) \in Q \times S\;|\;q \notin F_G \lor s \in F_{A_\psi}\}$ 
\State Solve safety game through greatest fixpoint equation $Y \mapsto Safe \cap Pre(Y)$.
\State \textbf{return} solution of the fixpoint equation
\end{algorithmic}
\end{algorithm}

\paragraph{Symbolic Product Construction.}
The CGS and the formula DFA are encoded as BDDs \cite{bryant_1986_graphbased}. Each structural component - states, actions, transition relation, and labelling function - is encoded as a BDD to enable scalable manipulation and fixpoint computation. These encodings are used in Algorithms~\ref{alg:symbolic-product} and 3 to symbolically construct the product graph and solve the safety game respectively following Def.~\ref{def:product-graph} and~\ref{def:safety-set-game}.
\begin{definition}[Product of CGS and DFA]
    \label{def:product-graph}
    Let $G = \langle Ag, AP, Q, q_0, \allowbreak {Act}_{a \in Ag},\delta,  F,\lambda\rangle$ be a CGS and $\mathcal{A}_\psi = \langle S, s_0, 2^{AP}, \Delta,  F' \rangle$ a DFA encoding the temporal formula $\psi$. The product graph $G \otimes \mathcal{A}_\psi$ for a coalition $A$ of agents is a structure whose set of states is $Q \times S$, with initial state $(q_0, s_0')$ for $s_0' = \Delta(s_0, \lambda(q_0))$. Transitions $\to$ in the product are defined as
    \begin{align*}
        & ((q, s), \alpha_A) \to (q',s') \iff \\
        & \exists \alpha_{-A}.(q, \alpha_A \cup \alpha_{-A}, q') \in \delta \land (s, \lambda(q'), s') \in \Delta
    \end{align*}
    where $\alpha_{-A}$ are the joint actions from the non-coalition agents.
\end{definition}
\begin{algorithm}
\caption{Symbolic Product Construction}
\label{alg:symbolic-product}
\begin{algorithmic}[1]
\Statex \textbf{Input:} CGS $G$, DFA $A_\psi$, Coalition $A \subseteq Ag$
\Statex \textbf{Output:} Initial Safety set $Safe \subseteq Q \times S$
\State $\delta_A(\vec{q}, \vec{a}_A, \vec{q'}) := \exists \vec{a}_{-A}.\; \delta(\vec{q}, \vec{a}_A, \vec{a}_{-A}, \vec{q'})$
\State $\delta'(\vec{q}, \vec{a}_A, \vec{q'}, \vec{s}, \vec{s'}) := \delta_A(\vec{q}, \vec{a}_A, \vec{q'}) \land \Delta(\vec{s}, \vec{q'}, \vec{s'})$
\State $q_0' :=\text{SwapVariables}(\exists \vec{s}.(q_0 \land \text{SwapVariables}(\Delta(\vec s , \vec{q'}, \vec{s'}), \vec{q'}, \vec{q})), \vec{s'}, \vec s)$
\State $R_0 := q_0'$; $R := \emptyset$
\Repeat
    \State $R \gets R \cup R_0$
    \State $R_0 \gets \text{Post}(R_0) \setminus R$
\Until{$R_0 = \emptyset$}
\State $Safe := R \land (\neg F \lor F')$
\State \Return $Safe$
\end{algorithmic}
\end{algorithm}

%
Algorithm~\ref{alg:symbolic-product} proceeds in three phases: (i) abstraction, (ii) composition, and (iii) pruning. The transition relation of the CGS is first abstracted w.r.t.~coalition \(A\), by existentially quantifying over non-coalition actions in line 1. This yields a BDD encoding all transitions that \(A\) can enforce. Next, the product transition relation is constructed by combining the abstracted CGS relation with the DFA's transition relation (line 2), synchronizing on propositional labellings. This is enabled by replacing the formulas over the propositions $AP$ in the DFA labels by the states in the CGS in which those propositional formulas hold. A symbolic reachability fixpoint starting from the initial state 
in line 3 then computes the set of reachable product states (lines 4-8). Finally, in line 9 the safety set is computed as described in Def.~\ref{def:safety-set-game}. This set is the domain of the safety game solved in the next stage.
\begin{definition}[Safety Set and Safety Game]
\label{def:safety-set-game}
    Let $P \subseteq Q \times S$ be the set of product states. The initial safety set is defined as \cite{belardinelli_2018_alternatingtime}: 
    \[
    Safe = \{(q, s) \in P\;|\; q \notin F \lor q \in F'\}
    \]
    A safety game is a tuple $(V, \Sigma, E, Safe)$, where $V$ are the vertices, $\Sigma$ the action alphabet, $E \subseteq V \times \Sigma \times V$ the transition relation, and $Safe \subseteq V$ the safety set \cite{gradel_2002_automata}. Solving the safety game consists of computing the greatest fixpoint operation $Y \mapsto Safe \cap Pre(Y)$ where $Pre(Y) = \{v\,|\, \exists \sigma.\forall v'. (v, \sigma, w) \in E \to w \in Safe\}$.
    Intuitively, this computes the set of states from which the coalition has a strategy to remain within $Safe$ indefinitely.
\end{definition}
\begin{algorithm}
\caption{Symbolic Fixpoint Solution}
\label{alg:symbolic-fixpoint}
\begin{algorithmic}[1]
\Statex \textbf{Input:} Product transition relation $\delta'$, initial safety set $Y$
\Statex \textbf{Output:} Set of states satisfying $\langle\!\langle A \rangle\!\rangle \psi$
\State $Y_0 \gets Y$
\Repeat
    \State $T_{bad}(\vec{q},\vec{s}, \vec{a}_A, \vec{q'}, \vec{s'}) \gets \delta' \land \neg Y_i$
    \State $B(\vec{q}, \vec s, \vec{a}_A) \gets \exists\, \vec{q'}, \vec{s'} .\; T_{\text{bad}}$
    \State $G(\vec q, \vec{s}, \vec{a}_A) \gets \neg B$
    \State $\text{Pre}(Y_i)(\vec{q}, \vec{s}) \gets \exists\, \vec{a}_A.\; G$
    \State $Y_{i+1} \gets Y_i \land \text{Pre}(Y_i)$
\Until{$Y_{i+1} = Y_i$}
\State \Return $Y_i$
\end{algorithmic}
\end{algorithm}

\paragraph{Symbolic Fixpoint Solution Algorithm.}
The symbolic fixpoint algorithm presented in Alg.~3 computes the controllable safety region for coalition $A$. At each iteration, any states in the current safety set from which unsafe transitions exist are identified and removed via a pre-image operator that quantifies over coalition actions. The fixpoint converges to the maximal set of product states from which $A$ can guarantee the satisfaction of $\psi$.
\begin{theorem}[Soundness of Alg.~\ref{alg:game-solving-finite}]
The symbolic procedure presented in Alg.~\ref{alg:game-solving-finite}
correctly computes the set of states satisfying $\langle\!\langle A\rangle\!\rangle \psi$ interpreted over finite-trace semantics.
\end{theorem}

This follows from Theorem 1 in \cite{belardinelli_2018_alternatingtime}, together with the fact that the binary encodings of the CGS and DFA as BDDs preserve their operational semantics (see Sec.~\ref{sec:symbolic-encodings} in the supplementary material), and the symbolic operations applied in Alg.~\ref{alg:symbolic-product} and \ref{alg:symbolic-fixpoint} are sound implementations of the corresponding explicit-state procedures.

\section{Parity-Game Symbolic Algorithm for Infinite Traces}
\label{sec:parity-alg}

In this section we present a symbolic reduction of ATL\(^*\) model checking on infinite traces to the solution of a parity game, which is a two-player game played on infinite traces on a graph. 

The ATL\(^*\) model checking algorithm again follows a recursive labelling approach, where the interesting cases are
strategy formulas \(\langle\!\langle A \rangle\!\rangle \psi\), where \(\psi\) is an LTL path formula. Intuitively, to verify these formulas we translate \(\psi\) into a deterministic parity automaton (DPA), encoded symbolically using BDDs. The corresponding parity game captures the strategic interactions between coalition \(A\) and the environment, with winning conditions derived from the DPA's acceptance condition. The complete procedure is shown in Alg.~\ref{alg:parity-infinite}.

\paragraph{Symbolic Parity Game Construction.}
Given a CGS and a coalition \(A\), we construct a two-player parity game described as follows:
\begin{definition}[Parity Game from CGS and DPA]
\label{def:parity-game-construction}
Given a CGS $G =  \langle Ag, AP,Q,{Act}_{a \in Ag},\delta, q_0, F,\lambda\rangle$ and a deterministic parity automaton (DPA) $A_\psi = \langle S, \Delta, s_0, \omega \rangle$, we define a parity game $\mathcal{G} = (V, E, \Omega)$ as follows:

\begin{itemize}
    \item The set of vertices is $V = V_0 \cup V_1$, where:
    \begin{itemize}
        \item $V_0 \subseteq Q \times S$ are positions controlled by player~0 (coalition $A$),
        \item $V_1 \subseteq Q \times S \times JAct_A$ are positions controlled by player~1, i.e., $Ag \setminus A$, where $JAct_A$ is the set of joint actions available to coalition $A$.
    \end{itemize}

    \item The edge relation $E = E_{0\to1} \cup E_{1\to0}$ is defined by:
    {\small
    \begin{eqnarray*}
    E_{0 \to 1} =\\
     \left\{ ((q, s), (q, s, a_A)) \;\middle|\; (q, s) \in V_0,\; a_A \in JAct_A \right\} \\
    E_{1 \to 0} =\\ 
    \begin{aligned}
    \{((q, s, a_A), (q', s')) \;\big|\;
    &\quad (s, \lambda(q'), s') \in \Delta \land\;\\ \exists a_{-A} \in JAct_{-A}.
    &\quad (q, a_A \cup a_{-A}, q') \in \delta\}
    \end{aligned}
    \end{eqnarray*}
    }
    \item The priority function $\Omega: V \to \mathbb{N}$ assigns to each vertex the priority of its automaton state:
    \[
    \Omega((q, s)) = \omega(s), \quad \Omega((q, s, a_A)) = \omega(s)
    \]
\end{itemize}
\end{definition}
The parity game is constructed over a symbolic state space that includes the current CGS state, the DPA state, and -- at player~1's vertices -- coalition $A$’s chosen joint action. This split encodes the alternating move structure required for parity games. Transitions from player~0 to player~1 copy the current state and append a coalition action; transitions from player~1 execute the CGS and DPA transitions synchronously.

\paragraph{Solving Parity Games.}
The winning region for player~0 in the resulting game corresponds to the set of CGS states from which coalition \(A\) can enforce the satisfaction of \(\psi\). The game can be solved using symbolic parity game solving algorithms such as the set-based progress measure algorithm from \cite{chatterjee_2018_quasipolynomial} to obtain a fully symbolic ATL\(^*\) model checking procedure.
\begin{algorithm}
    \caption{ATL\(^*\) Model Checking Symbolic Algorithm through Parity Games}
    \label{alg:parity-infinite}
    \begin{algorithmic}[1]
        \Statex \textbf{Input:} Symbolic CGS, ATL\(^*\) formula $\langle\!\langle A \rangle\!\rangle \psi$
        \Statex \textbf{Output:} Set $W$ of states satisfying the formula.
        \State Transform $\psi$ into symbolic DPA $A_\psi = \langle S, \Delta,s_0, \Omega \rangle$
        \State Compute reachable product states: \(Reach(\vec{q}, \vec{s})\)
        \State \(V_0(\vec{q}, \vec{s}) = Reach(\vec{q}, \vec{s})\)
        \State \(V_1(\vec{q}, \vec{s}, \vec{a_A}) = V_0 \land Act_A(\vec{q}, \vec{a_A})\)
        \State $E_{0\to1} = V_0(\vec q, \vec s) \land Id(\vec{q}, \vec{q'}) \land Id(\vec{s}, \vec{s'}) \land Act_A(\vec{q'}, \vec{a_A'})$
        \State
        $E_{1\to0} = V_1 \land \delta(\vec{q}, \vec{a_A}, \vec{a}_{-A}, \vec{q'}) \land \Delta(\vec{s}, \vec q, \vec{s'})$
        \State Construct game \(P = (V_0, V_1, E = E_{0\to1} \cup E_{1\to0}, \Omega)\)
        \State Solve parity game \(P\)
        \State \Return Winning region \(W\) for Player 0
    \end{algorithmic}
\end{algorithm}
\begin{theorem}[Soundness of Alg.~\ref{alg:parity-infinite}] \label{theo:alg:parity-infinite}
The symbolic procedure presented in Alg.~\ref{alg:parity-infinite}
correctly computes the set of states satisfying $\langle\!\langle A\rangle\!\rangle \psi$ interpreted over infinite-trace semantics.
\end{theorem}

Theorem~\ref{theo:alg:parity-infinite}
follows from the correctness of the symbolic encoding and the general result in \cite{alfaro_2002_from}, which shows that control problems with \(\omega\)-regular objectives (including LTL formulas) can be solved via parity games. 
By the correct encoding of CGS \( G \) and DPA \( \mathcal{A}_\psi \) as BDDs (see Sec.~\ref{sec:symbolic-encodings} in the supplementary material), and constructing the parity game as defined in Def.~\ref{def:parity-game-construction}, we obtain a sound symbolic reduction. The BDD-based operations faithfully implement the semantics of the parity game, and the computed winning region characterizes the set of states from which coalition \( A \) can enforce \( \psi \).

\section{Implementation}

\label{sec:implementation}
\begin{figure*}[ht]
    \centering
    \includegraphics[width=0.85\textwidth]{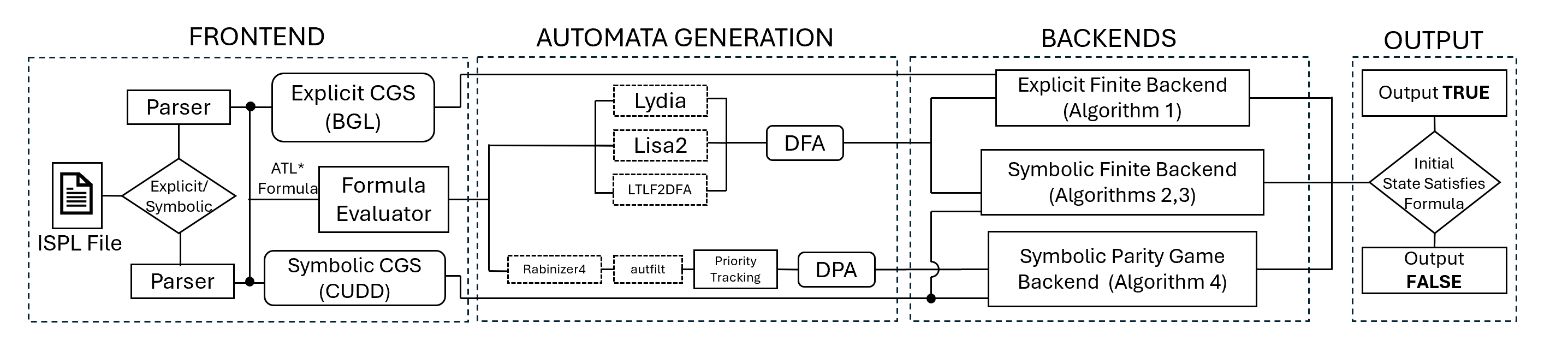}
    \caption{Overview of ATL\(^*\)AS tool architecture, showing the modular design of the frontend, automata generation pipeline, and solver backends. Dashed rectangles indicate the external components used.}
    \label{fig:tool-architecture}
\end{figure*}

We implemented the algorithms presented in Sec.~\ref{sec:finite-algs} and~\ref{sec:parity-alg} in our verification tool, ATL\(^*\)AS, which supports both finite- and infinite-trace semantics of ATL\(^*\). The repository for this tool can be accessed at \url{https://gitlab.doc.ic.ac.uk/sg2922/atlf_model_checker/-/tree/master?ref_type=heads}. The tool integrates three model checking backends: an explicit-state and a symbolic-state solver for ATL\(^*_f\), and a symbolic parity-game-based solver for ATL\(^*\).
The overall architecture of the tool is illustrated in Fig.~\ref{fig:tool-architecture}, highlighting the main components and their interactions across (i) the frontend, (ii) automata generation pipeline, and (iii) backend solvers.
\paragraph{Input Language.} ATL\(^*\)AS uses a subset of the ISPL language \cite{lomuscio_2015_mcmas}, extended to support the declaration of final states. The parser has been extended to support the full syntax of ATL\(^*\).

\paragraph{Explicit and Symbolic Backends.} For the explicit algorithms, transitions and state sets are implemented using adjacency lists and dynamic bitsets from the Boost Graph Library~\cite{bgl}. Symbolic representations use the CUDD package~\cite{a2025_cudd} to manipulate BDDs. Both explicit and symbolic pipelines share a common preprocessing stage that handles the LTL\(_f\)-to-DFA conversion described next.
\paragraph{Automata Translation for Finite Traces.} Translation of LTL\(_f\) path formulas into deterministic finite automata (DFA) is a critical bottleneck in ATL\(_f^*\) model checking. To mitigate this, ATL\(^*\)AS adopts a parallel translation strategy that exploits the complementary performance of existing tools across varied formula structures. Specifically, we run translators Lisa2 \cite{bansal_2024_dagbased}, Lydia \cite{giuseppedegiacomo_2021_compositional}, and LTLF2DFA \cite{whitemech_2021_github} concurrently as subprocesses. The first tool to produce a valid DFA terminates the others, and its output is used in the subsequent verification procedure. This asynchronous execution model significantly reduces automaton generation in practice.

\paragraph{Automata Translation for Infinite Traces.} In the infinite-trace setting, LTL path formulas are translated into deterministic parity automata (DPA) using Rabinizer4 \cite{ketnsk_2018_rabinizer} which implements the Safra-less translation through limit-deterministic Büchi automata (LDBA) \cite{esparza_2017_from}.  Rabinizer4 outputs a transition-based DPA, where priorities are assigned to transitions. We then translate it into a state-based DPA in order to build the parity game.
\paragraph{Parity Game Solving Backend.}
ATL\(^*\)AS integrates the symbolic, set-based algorithm from \cite{chatterjee_2018_quasipolynomial} to solve parity games in the infinite-trace ATL\(^*\) setting. This algorithm achieves quasi-polynomial complexity in symbolic operations and linear symbolic space, enabling efficient, fully symbolic game solving.

\section{Evaluation and Results}
\label{sec:evaluation}
We evaluate the performance and scalability of ATL$^*$AS
using a synthetic benchmark and a real-world cybersecurity scenario. The synthetic benchmark is based on a multi-agent counter system with two parameters: the maximum counter value \(C\) and the maximum number of steps \(S\) in the finite-trace case. This is inspired by the single-counter LTL\(_f\) synthesis benchmark suite \cite{whitemech_2021_benchmark}, but adapted to strategic reasoning. On the synthetic benchmark, two experiments were conducted: one scaling the size of the CGS and another scaling the size of the input ATL\(^*\) formula. Since there are no specific ATL\(^*\) model checking tools, we compare the parity-game-based ATL\(^*\)AS model checker against the MCMAS-SL[1G] tool on the fair scheduler synthesis benchmark. Finally, we assessed the symbolic ATL\(^*_f\) algorithm on a cybersecurity scenario to evaluate its practical applicability in a real-world multi-agent setting.

All experiments were run on an Intel Core™ i7-10700 CPU (2.90GHz) with 16GB RAM and SSD storage.

\subsection{Evaluation on Finite Traces}
\begin{figure}[t]
    \centering
    \includegraphics[width=0.9\linewidth]{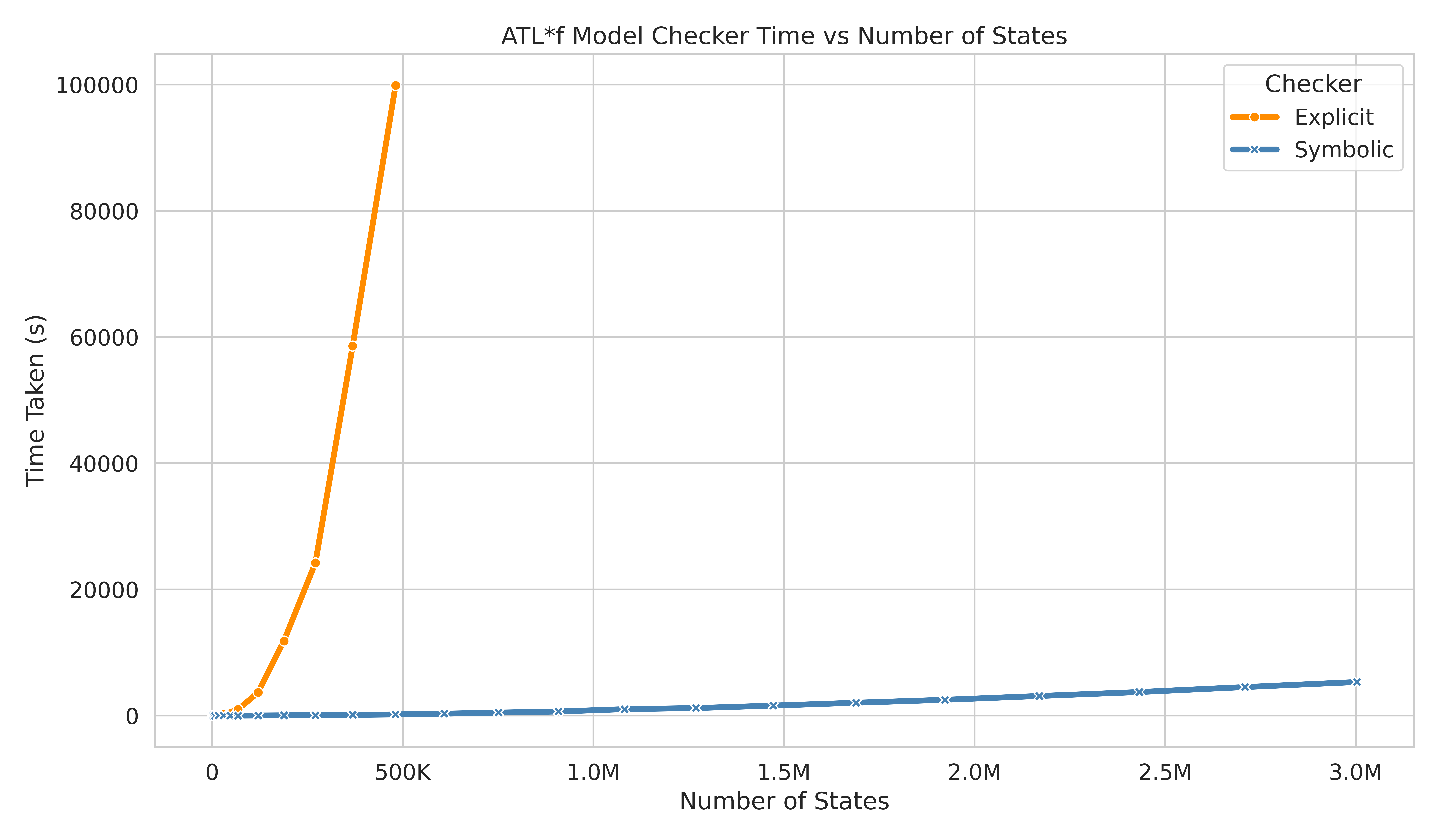}
    \caption{Exp.~1 -- Comparison of runtime performance between the explicit and symbolic model checkers.}
    \label{fig:finite-comparison-runtime}
\end{figure}

\begin{figure}[t]
    \centering
    \includegraphics[width=1\linewidth]{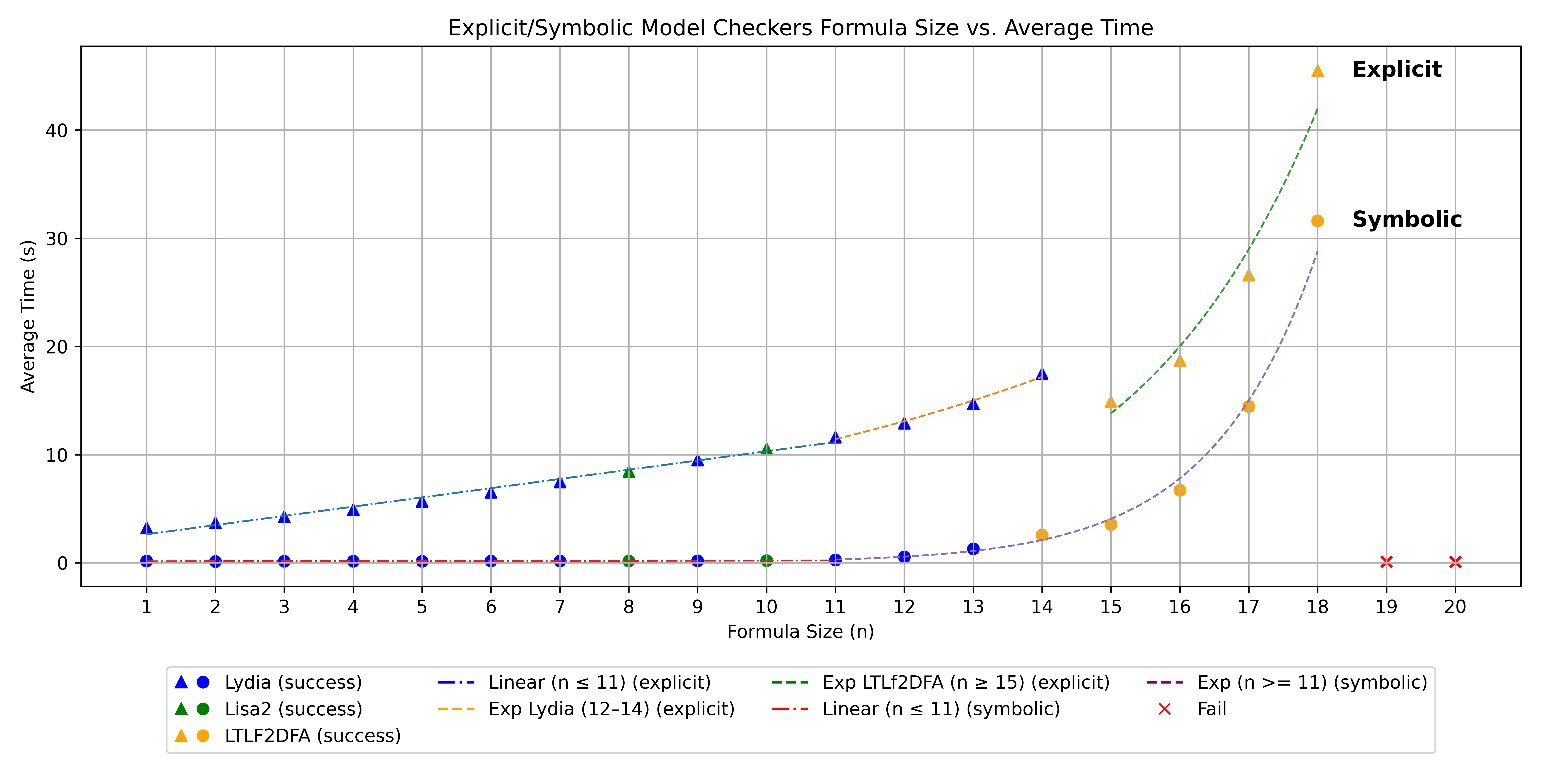}
    \caption{Exp.~2 -- Runtime of the explicit and symbolic model checkers as formula size increases. Marker colour indicates the tool used for DFA generation. Dashed lines show fitted linear and exponential curves.}
    \label{fig:finite-comparison-formula}
\end{figure}

\paragraph{Experiment 1: Scaling the CGS.}
We varied the benchmark parameters \(C\) and \(S\) to generate a range of models. The formula to verify is
$\langle\!\langle A \rangle\!\rangle \, \mathsf{F} \, \texttt{counter\_max}$,
where \texttt{counter\_max} holds in states where the counter reaches its maximum value $C$. Figure~\ref{fig:finite-comparison-runtime} shows the average runtime over five runs for each instance.
The explicit checker exhibits a steep increase in runtime, with performance degrading significantly beyond 100k states. This is likely due to the cost of enumerating the full state-space product and executing repeated fixpoint computations. For the largest instance (nearly 500k states, $C=800$, $S=800$), verification required over 27 hours.
In contrast, the symbolic checker scales more efficiently. While still showing polynomial growth (approximately quadratic), runtimes remain far lower. Even with over 3M states ($C = 2000$, $S = 2000$), the symbolic checker completes in under 1.5 hours, demonstrating significantly better scalability.
\paragraph{Experiment 2: Scaling the formula size.}
To assess the impact of temporal formula complexity, we fixed the CGS parameters to \(C = 40\), \(S = 35\), yielding approximately 1k states, and varied the depth of a nested formula of the form:
\[
\langle\!\langle A, B \rangle\!\rangle \, \mathsf{F} \, p_1 \land \mathsf{X}(\mathsf{F} \, p_2 \land \mathsf{X}(\mathsf{F} \, p_3 \land \ldots \land \mathsf{X}(\mathsf{F} p_n)))
\]
where each \(p_i\) holds in states where the counter equals \(i\). Formula size \(n\) ranges from 1 to 20. Figure~\ref{fig:finite-comparison-formula} reports average runtimes and indicates the tool used for DFA construction.

Both the explicit and symbolic implementation scale linearly up to \(n = 11\), with the symbolic version consistently faster due to more efficient product construction and fixpoint solving. For \(n \geq 12\), exponential growth emerges, dominated by the cost of automaton translation, although the symbolic checker remains more efficient, maintaining a practical runtime even for the largest successful instance of \(n = 18\).

For \(n \geq 19\), all automata generation tools fail due to resource exhaustion. The parallelised translation strategy, which selects Lydia and Lisa2 for small formulas and LTLF2DFA for larger ones, proves critical. Without this, failure would occur at \(n = 17\). This hybrid approach improves robustness across the range of formula sizes.

\paragraph{Discussion.}
The experiments demonstrate our symbolic model checking algorithm's superior scalability and efficiency over the explicit implementation across both dimensions of the problem: system size and formula complexity. The modular use of multiple translation tools further enhances reliability and performance.
\subsection{Finite Traces vs. Infinite Traces Comparison}
\begin{figure}[t]
    \centering
    \includegraphics[width=0.8\linewidth]{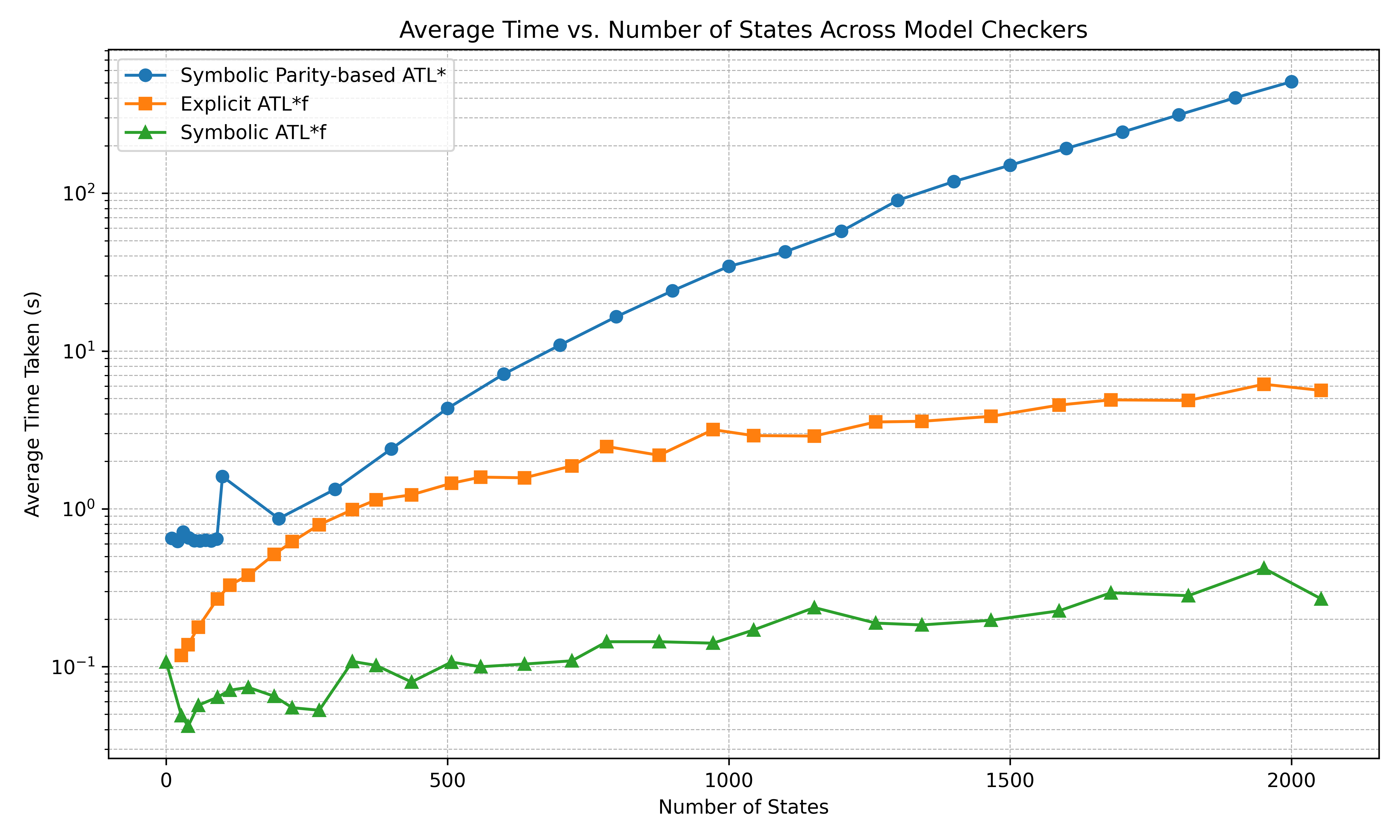}
    \caption{Exp.~1: Comparison of runtime versus model size for finite-trace (explicit and symbolic) and infinite-trace (parity-game–based) model checkers.}
    \label{fig:infinite-finite-comp-exp1}
\end{figure}

\begin{figure}[t]
    \centering
    \includegraphics[width=0.8\linewidth]{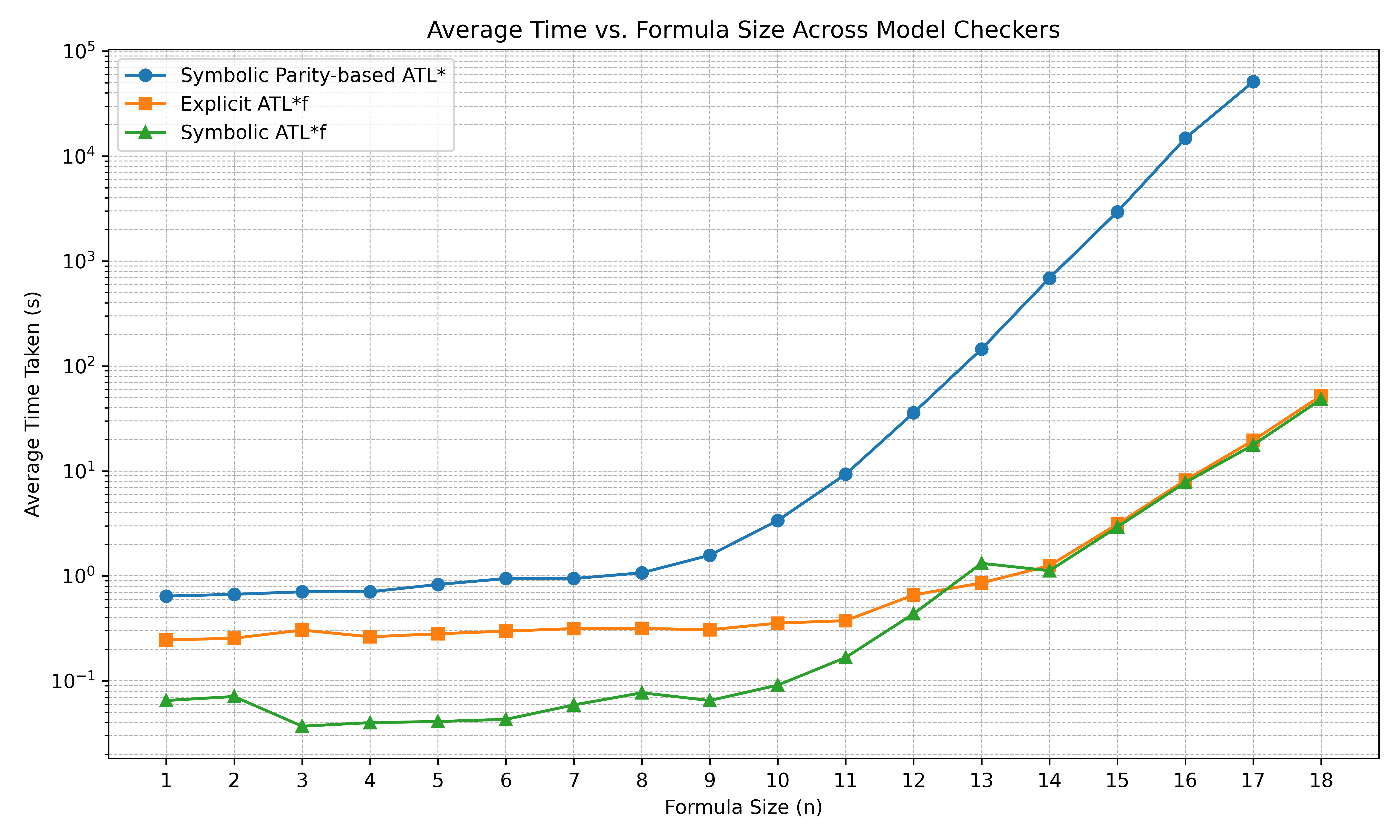}
    \caption{Exp.~2: Comparison of runtime versus formula size for finite-trace (explicit and symbolic) and infinite-trace (parity-game–based) model checkers.}
    \label{fig:infinite-finite-comp-exp2}
\end{figure}

Figures~\ref{fig:infinite-finite-comp-exp1} and~\ref{fig:infinite-finite-comp-exp2} compare the performance of all model checking algorithms -- explicit finite-trace, symbolic finite-trace, and parity-based infinite-trace -- on Experiments 1 and 2, respectively.

In both experiments, the finite-trace model checkers, particularly the symbolic variant, significantly outperform the parity-based infinite-trace approach, often by several orders of magnitude. This gap is especially pronounced in the formula-scaling experiment, since in practice, LTL\(_f\)-to-DFA conversions often yield automata far smaller than their worst-case bounds, while the corresponding \(\omega\)-automata for LTL can remain large even when generated with optimized tools like Rabinizer4. These results confirm the hypothesis presented in \cite{belardinelli_2018_alternatingtime} that when the system being modelled allows a finite-trace representation, ATL\(^*_f\) model checking is substantially more tractable. Obviously, modelling system executions as finite traces is not always appropriate/possible.


\subsection{Comparison with MCMAS-SL[1G]}

While no existing tools implement model checking specifically for ATL\(^*\), verification can be performed indirectly using logics that strictly subsume it. One such logic is the one-goal fragment of Strategy Logic (SL[1G]), whose syntax and semantics generalise those of ATL\(^*\), and whose model checking problem is also 2EXPTIME-complete~\cite{mogavero_2014_reasoning}. This makes SL[1G] a natural reference point for comparison, particularly via the SL[1G] model checking extension of the MCMAS model checker~\cite{cermk_2014_sl1g}.

To compare ATL$^*$AS to MCMAS-SL[1G], we use the fair scheduler synthesis benchmark in \cite{cermk_2014_sl1g}, parameterised by the number of processes \(n\) competing for access to a shared resource. The property to be verified ensures fairness for each process: if a process is waiting (atom $wt_i$), it will eventually cease waiting (by acquiring access or giving up).
This property can be encoded in ATL\(^*\) as:
\[
\bigwedge_{i=1}^n \langle\!\langle P_i\rangle\!\rangle\, \mathsf{G} (wt_i \rightarrow \mathsf{F} \neg wt_i)
\]

Since SL[1G] subsumes ATL\(^*\) this property can also be expressed in SL[1G]. The runtime comparison is shown in Table~\ref{tab:sl1g_comp_results}.
\begin{table}[H]
\centering
\caption{Runtime comparison between MCMAS-SL[1G] and ATL\(^*\)AS as the number of processes increases.}
\label{tab:sl1g_comp_results}
\resizebox{\columnwidth}{!}{%
\begin{tabular}{|c|c|c|c|}
\hline
\textbf{\#Processes} & \textbf{\#States} & \textbf{ATL\(^*\)AS Runtime (s)} & \textbf{MCMAS-SL[1G] Runtime (s)} \\
\hline
2 & 9 & 0.94 & 0.15 \\
3 & 21 & 1.63 & 3.96 \\
4 & 49 & 3.17 & 234.25 \\
5 & 113 & 6.85 & 4339.47 \\
6 & 257 & 21.40 & 60851.20 \\
\hline
\end{tabular}%
}
\end{table}
ATL\(^*\)AS
outperforms MCMAS-SL[1G] consistently from \(n = 3\) onwards, with the performance gap widening exponentially. For \(n = 2\), MCMAS-SL[1G] is marginally faster, likely due to lower fixed overheads for small models. However, as model size increases, ATL\(^*\)AS achieves runtimes up to four orders of magnitude faster.
Two factors contribute to this advantage. First, ATL\(^*\)AS leverages modern external tools for automata generation. Specifically, the DPA is produced using a Safra-less translation based on LDBA~\cite{esparza_2017_from}, which yields smaller automata than the Generalised Büchi Automata (GBA)-based approach used in MCMAS-SL[1G]. Second, for game solving, ATL\(^*\)AS employs symbolic quasi-polynomial parity game algorithms, while MCMAS-SL[1G] relies on a symbolic implementation of Zielonka’s algorithm \cite{zielonka_1998_infinite}. These combined improvements in algorithmic and architectural design, along with a simpler logic, enable ATL\(^*\)AS to achieve significantly better scalability in practice.

\subsection{Cybersecurity Modelling Scenario}
\label{sec:modelling}
To demonstrate the applicability of $\text{ATL}_f^*$ verification in real-world contexts, we model a multi-agent cybersecurity scenario inspired by the CyMARL simulation environment~\cite{wiebe_2023_learning}. 

\paragraph{Scenario Description.}
The system consists of five critical servers, protected by two defenders and targeted by a single intelligent attacker. Each server represents a critical service within an enterprise infrastructure. Defenders act as automated intrusion prevention systems (IPSs) constrained by a shared budget, while the attacker models an advanced persistent threat (APT) executing targeted campaigns to compromise one of three objectives:
\begin{itemize}
    \setlength\itemsep{0em}
    \item \textbf{Confidentiality Breach}: gain administrator-level access to exfiltrate sensitive data.
    \item \textbf{Integrity Violation}: insert or modify files to corrupt system behaviour.
    \item \textbf{Availability Attack}: deploy denial-of-service malware to render a host unusable.
\end{itemize}
Each defender action incurs a cost that reflects its real-world impact on system performance. To preserve decidability while capturing the defenders' partial observability, action availability is restricted dynamically via heuristic protocols that determine what responses are permissible based on observed evidence.

The model is finite-horizon, reflecting fixed-duration attacks, and is formalised as a Concurrent Game Structure (CGS) with final states suitable for ATL\(^*_f\) verification.

\paragraph{State Space and Observables.}
Each global state includes:
\begin{itemize}
  \setlength\itemsep{0em}
  \item \textbf{Attacker position} \(p \in \{0,\dots,4\}\): the server on which the attacker currently resides.
  \item \textbf{Privilege levels} \(\text{priv}[i]\in\{0,1,2\}\) for each server \(i\): 0 = no access, 1 = user, 2 = admin.
  \item \textbf{Vulnerability map} \(\text{scanned}[i]\in\{0,1\}\): whether server \(i\) has been scanned.
  \item \textbf{Availability map} \(\text{down}[i]\in\{0,1\}\): whether server \(i\) is unavailable due to a DoS payload.
  \item \textbf{Defender budgets} \(B\): remaining combined budget for defenders \(D_{1}\) and \(D_{2}\).
  \item \textbf{Suspicion buckets} \(\sigma_{j,i}\in\{0,1,2\}\): current suspicion level that defender \(D_{j}\) assigns to server \(i\).
  \item \textbf{Step counter} \(t\): the current time step, \(0 \le t \le T\).
  \item \textbf{Server flags} \(F_{i} = (F_{i,1},\dots,F_{i,6}) \in \{0,1\}^6\): six Boolean indicators per server \(i\).
\end{itemize}
When the defender performs a \(\texttt{monitor}(i)\) action on a server they observe the set of flags describing what events have ocurred on that server. We refer to \cite{wiebe_2023_learning} for full details on this scenario.

\paragraph{Agent Actions.}
At each timestep, the attacker and both defenders each choose a single action. Logical preconditions (e.g., scanning before exploit) and access constraints are enforced by the protocol function.
\paragraph{Defender actions}  
\begin{itemize}
  \setlength\itemsep{0em}
  \item \texttt{monitor}(\(i\)): query server \(i\) to observe its flag vector \(F_i\).
  \item \texttt{remove}(\(i\)): evict any user‐level malicious process on \(i\).
  \item \texttt{restore}(\(i\)): re‐image \(i\), clearing all sessions and processes.
  \item \texttt{do\_nothing}: skip action.
\end{itemize}
In the \emph{integrity} scenario, defenders also have an \texttt{analyze}(\(i\)) action to inspect file‐system metadata to detect tampering and a \texttt{data\_repair}(\(i\)) action to remove or revert modified files on \(i\).
\paragraph{Attacker actions}  
\begin{itemize}
  \setlength\itemsep{0em}
  \item \texttt{scan}(\(i\)): probe \(i\) for vulnerabilities.
  \item \texttt{exploit}(\(i\)): leverage a known vulnerability to obtain user‐level access.
  \item \texttt{escalate}(\(i\)): promote from user to admin privileges on \(i\).
  \item \texttt{do\_nothing}: skip action.
\end{itemize}
Additionally in the \emph{integrity} scenario: \texttt{tamper}(\(i\)), modifying or corrupting files (requires appropriate privileges) and in the \emph{availability} scenario: \texttt{deny}(\(i\)), install a DoS‐style payload to render \(i\) unavailable.

\paragraph{Imperfect Information Abstraction.}
While ATL\(^*\) assumes perfect information, defenders in this scenario must act under partial observability. Since model checking ATL\(^*\) with imperfect information is undecidable under perfect recall strategies \cite{raphalberthon_2017_decidability}, we approximate this setting with the perfect-information framework using a dynamic action restriction protocol.

Each defender maintains a per-server suspicion bucket \(\sigma_{j,i} \in \{0, 1, 2\}\) which is updated based on observed flags after a \(\texttt{monitor}\) action. The bucket determines which defensive actions \(D_j\) may invoke on \(i\). This mechanism prevents defenders from unrealistically reacting to attacker actions they cannot observe.

\paragraph{Suspicion Heuristics.}
To compute suspicion levels from observed flags \(F_i\), we define four heuristics reflecting different defensive philosophies.
\begin{enumerate}
    \item \textbf{Conservative} (risk weighted): each flag \(F_{i,k}\in\{0,1\}\) carries a danger weight \(w_k\) (e.g., higher for exploit or admin process flags).  Defender \(D_j\) computes the weighted sum \(S_i \;=\;\sum_{k=1}^6 w_k\,F_{i,k}\), and applies thresholds \(0\le T_1 < T_2\) so that \(\sigma_{j,i}=0\) if \(S_i<T_1\), \(\sigma_{j,i}=1\) if \(T_1\le S_i<T_2\), and \(\sigma_{j,i}=2\) otherwise.
    \item \textbf{Aggressive} (zero-trust): immediate escalation to \(\sigma = 2\) if critical flags are seen (e.g., admin process).
    \item \textbf{Proportional}: normalised version of the conservative heuristic, following percentile-based risk matrices approaches.
    \item \textbf{Diversity}: suspicion based on the number of flag types seen, inspired by the MITRE cyber kill-chain phases \cite{lockheedmartin_2025_cyber}.
\end{enumerate}

\begin{figure}[H]
    \centering
    \includegraphics[width=0.9\linewidth]{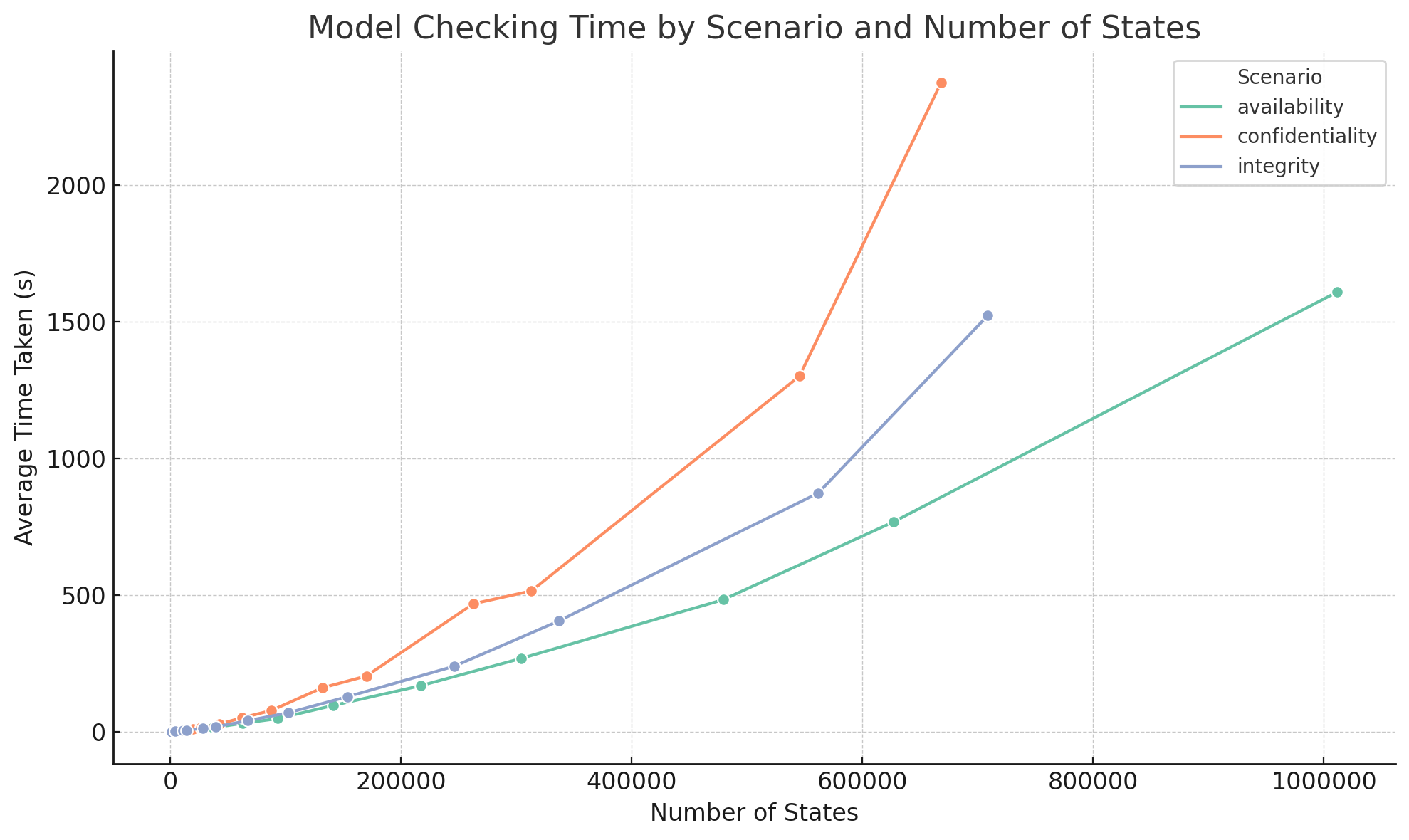}
    \caption{Runtime of symbolic model checker for different attacker goals as number of states increases.}
    \label{fig:modelling-results}
\end{figure}

\paragraph{Evaluation.}
Using the symbolic finite-trace version of ATL$^*$AS, we evaluate verification performance across three attacker objectives (confidentiality breach, integrity violation, availability attack), as the model size increases (via time steps $n$ and defender budget $b$). The formula used in this verification is $\langle\!\langle \text{Attacker} \rangle\!\rangle \mathsf F \mathsf G\; \text{compromised\_1}$. Figure~\ref{fig:modelling-results} shows that verification remains tractable even for models with over one million states, completing in under one hour.

We also verify strategic properties over defender coalitions. For instance, we compute the minimal defender budget required to ensure that no server remains compromised in the confidentiality scenario after 10 steps. The property is encoded as:
\[
\langle \!\langle D_1, D_2\rangle\!\rangle \mathsf F \mathsf G \neg \text{compromised\_1}
\]

Using binary search over $b \in [1..20]$, we find the minimal budget for which the formula holds under four defender heuristics, and thus also the most cost-effective strategy for the defenders in this model.
\begin{table}[H]
\centering
\begin{tabular}{c|c}
\textbf{Heuristic} & \textbf{Minimum Budget} \\
\hline
Proportional & 10 \\
Aggressive & 12 \\
Conservative & 14 \\
Diversity & 14
\end{tabular}
\caption{Minimum budget required to guarantee defence under each heuristic.}
\label{tab:min-budget}
\end{table}
These results show how ATL\(^*_f\) model checking can offer formal guarantees on the effectiveness of defensive strategies, revealing optimal operational thresholds under uncertainty. 

\section{Conclusions}
In this paper we introduced symbolic algorithms for model checking strategy logics ATL\(^*_f\) (Sec.~\ref{sec:finite-algs}) and ATL\(^*\) (Sec.~\ref{sec:parity-alg}) over finite and infinite traces respectively. 
in the ATL$^*$AS model checker, alongside an explicit ATL\(^*_f\) baseline (Sec.~\ref{sec:implementation}). Experimental results (Sec.~\ref{sec:evaluation}) demonstrate that the symbolic ATL\(^*_f\) algorithm achieves significant performance gains over its explicit counterpart, and that the symbolic ATL\(^*\) model checker outperforms the current state-of-the-art MCMAS-SL[1G] tool despite their shared 2EXPTIME complexity. The evaluation further confirms that finite-trace verification is considerably more efficient than infinite-trace approaches due to simpler automata constructions and model checking procedures. 
The tool’s applicability was demonstrated through a 
cybersecurity scenario, showcasing its utility in verifying attacker-defender strategies in complex multi-agent systems (Sec.~\ref{sec:modelling}).

\paragraph{Acknowledgments.} The research described in this paper was partially supported by the EPSRC (grant number EP/X015823/1) and by the Moro-Barry family.


\bibliographystyle{plain} \bibliography{references}

\clearpage
\appendix
\section{Symbolic Encodings}
\label{sec:symbolic-encodings}
We detail the symbolic encodings used in ATL$^*$AS for Concurrent Game Structures (CGSs), Deterministic Finite Automata (DFAs), and Deterministic Parity Automata (DPAs). These encodings ensure that all symbolic manipulations used in the algorithms correctly represent their respective structures, supporting the soundness claims in Sections~\ref{sec:finite-algs} and~\ref{sec:parity-alg}.
\subsection{Symbolic Encoding of CGSs}
Each component of the CGS is encoded using Binary Decision Diagrams (BDDs), enabling compact representation and efficient manipulation:
\paragraph{States.} Let $Q$ be the set of states with $|Q| = N$. Each state $s_i \in Q$ is encoded as a binary vector $\vec{q} = (q_0, \dots, q_{n-1})$ with $n = \lceil \log_2 N \rceil$. Transitions use a second vector $\vec{q'}$ to represent successor states.
\paragraph{Actions.} For $m$ agents $A_1, \dots, A_m$, each with action domain $\text{Act}_i$, we introduce Boolean vectors $\vec{a_i}$ of length $n_i = \lceil \log_2 |\text{Act}_i| \rceil$. The global joint action is encoded as $\vec{a} = \bigcup_i \vec{a_i}$.
\paragraph{Transition Relation.} Each transition $(s, \alpha, s') \in \delta$ is encoded as a conjunction over $(\vec{q}, \vec{a}, \vec{q'})$. The full transition relation is represented by the BDD:
\[
    \delta(\vec{q}, \vec{a}, \vec{q'}) = \bigvee_{(s, \alpha, s') \in \delta} [\text{enc}_\text{c}(s) \land \text{enc}_\text{a}(\alpha) \land \text{enc}_\text{n}(s')]
\]
\paragraph{Labelling Function.} The atomic proposition labelling $\lambda: Q \to 2^{AP}$ is inverted to $\lambda': AP \to \mathcal{B}(\vec{q})$, mapping each proposition $p$ to a BDD over the states in which $p$ holds.
\paragraph{Initial and Final States.} The initial state $q_0$ is a BDD over $\vec{q}$, and the set of final states $F \subseteq Q$ is encoded as a BDD over $\vec{q}$.
\subsection{Symbolic Encoding of DFAs}

\paragraph{State Variables.} For $M$ DFA states, use $\vec{d} = (d_0, \dots, d_{m-1})$ and $\vec{d'}$ with $m = \lceil \log_2 M \rceil$ to encode current and next DFA states.

\paragraph{Transition Conditions.} Each DFA transition $d \xrightarrow{\phi} d'$ is labelled with a propositional formula $\phi$ in DNF. Literals $p$ and $\neg p$ are replaced by $\lambda'(p)$ and $\neg \lambda'(p)$ respectively, yielding a BDD over next-state CGS variables $\vec{q'}$ via a \texttt{VectorCompose} from $\vec{q}$ to $\vec{q'}$.

\paragraph{Transition Relation.} Each transition becomes a conjunction over $\vec{d}$, $\vec{q'}$, and $\vec{d'}$. The complete DFA transition relation is a BDD $\Delta(\vec{d}, \vec{q'}, \vec{d'})$.

\subsection{Symbolic Encoding of DPAs}
DPAs are encoded symbolically as follows:
\paragraph{Structure.} Let $A = (S, \Sigma, \Delta, s_0, \Omega)$ be a state-based DPA with $k$ priority levels. Use $\vec{s}, \vec{s'}$ for current and next states, $\vec{p}$ for input propositions, and $\vec{c} = (c_0, \dots, c_{\ell-1})$ with $\ell = \lceil \log_2 k \rceil$ for priorities.

\paragraph{Transition Relation.} Represented as a BDD $\Delta(\vec{s}, \vec{p}, \vec{s'})$.
\paragraph{Priority Function.} Encoded as $\Omega(\vec{s}, \vec{c})$, associating each state with its parity index.
\section{Implementation Details}
\subsection{Handling Transition-Based Parity Acceptance}
Rabinizer4 produces DPA with transition-based acceptance, i.e. priorities are assigned to transitions rather than states. However, symbolic parity game construction used in ATL\(^*\)AS requires state-based priority assignments to correctly define the priority function $\Omega : V \to \mathbb{N}$ over the game's vertices.
To bridge this gap, we post-process the automaton using Spot's \texttt{autfilt} tool \cite{duretlutz_2016_spot}, which attempts to transform the automaton into an equivalent DPA with state-based parity acceptance.
\paragraph{Fallback Priority Encoding.} In some cases, \texttt{autfilt} fails to produce a state-based DPA and instead returns an automaton qith a more expressive acceptance condition (e.g. Rabin or Büchi). To handle these cases, we implement a custom transformation inspired by the priority tracking method of \cite{schewe_2014_determinising}. This procedure performs a breadth-first traversal of the DPA and assigns to each reachable state the priority of the incoming transition. The resulting structure is a valid state-based DPA and enables uniform construction of the parity game across all LTL formulas.

\section{Benchmark Description}
The benchmark used in our evaluation is a parameterised multi-agent counter system, designed to assess scalability across both finite- and infinite-trace variants of ATL$^*$. It extends the single-agent counter system from the LTL\(_f\) synthesis benchmark suite~\cite{whitemech_2021_benchmark} to a two-agent concurrent game structure (CGS) suitable for strategic reasoning.

The system models two agents, A and B, each capable of incrementing a shared counter (\texttt{i}) or waiting (\texttt{w}). For the finite-trace experiments, the model is parameterised by the counter bound \(C\) and the maximum number of steps \(S\), which together determine the size and structure of the CGS. In the infinite-trace setting, only the counter bound \(C\) is used, and the step count is unbounded as the system evolves indefinitely.

\begin{figure}[H]
    \centering
    \includegraphics[width=0.8\linewidth]{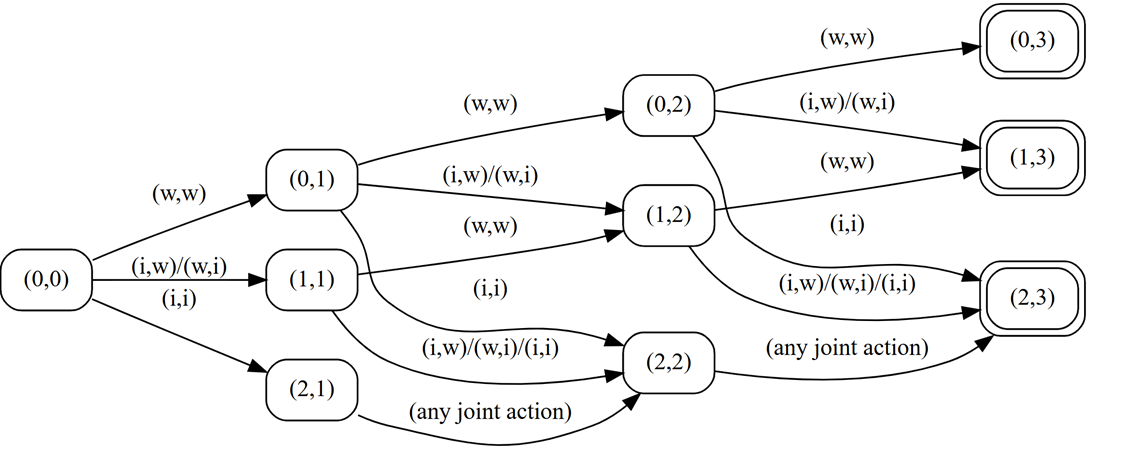}
    \caption{Example CGS instance from the benchmark with C=2,S=3}
    \label{fig:counter-cgs}
\end{figure}

An example instance from the finite-trace benchmark is shown in Figure~\ref{fig:counter-cgs}, where each state is labelled with its \((\texttt{curr\_count}, \texttt{step})\) value, and transitions are labelled with the joint actions of the agents.

\section{Rabin-Tree Automata Approach for ATL\(^*\) Model Checking}
As a comparison to the symbolic parity-game-based procedure presented in this paper, we implemented the classical approach for ATL\(^*\) model checking as described by Alur, Henzinger and Kupferman \cite{alur_2002_alternatingtime}. This algorithm verifies whether $G, q \models \langle\!\langle A\rangle\!\rangle \psi$ in four main steps:
\begin{enumerate}
    \item \textbf{Execution-tree automaton}. Construct a nondeterministic Büchi tree automaton $\mathcal{A}_{G,q,A}$ whose accepted trees are exactly the execution trees rooted at $q$ under all strategies of $A$.
    \item \textbf{Specification automaton}. Translate the LTL formula $\psi$ into a Deterministic Rabin Tree Automaton (DRTA) $A_\psi$ that accepts precisely those infinite paths satisfying $\psi$. 
    \item \textbf{Product NRTA}. Form the synchronous product
    \[
    \mathcal{A}_\text{prod} = \mathcal{A}_{G,q,A} \otimes \mathcal{A}_\psi
    \]
    yielding a nondeterministic Rabin tree automaton whose language is non-empty exactly when $A$ can enforce $\psi$ from $q$.
    \item \textbf{Decide emptiness of $\mathcal{A}_\text{prod}$} via the Kupferman-Vardi algorithm \cite{kupferman_1998_weak}. $L(\mathcal{A}_\text{prod}) \neq \emptyset$, then $q \models \langle\!\langle A\rangle\!\rangle \psi$. 
\end{enumerate}
We implemented this in a hybrid manner, performing steps 1-3 symbolically and step 4 explicitly. For Step~2 we again used the tool Rabinizer4 \cite{ketnsk_2018_rabinizer} in this construction, since it employs a Safra-less method that generates smaller DRAs than classic approaches. We then apply a lifting construction to obtain a tree automaton from the word automaton generated by the tool.

For the non-emptiness check in Step~4, we construct a Weak Rabin Alternating Automaton from the NRTA, apply the construction in \cite{kupferman_1998_weak} to generate a Weak Alternating Automaton and use the algorithms defined in \cite{kupferman_2000_an} to solve its non-emptiness. 

\begin{figure}[H]
    \centering
    \includegraphics[width=0.75\linewidth]{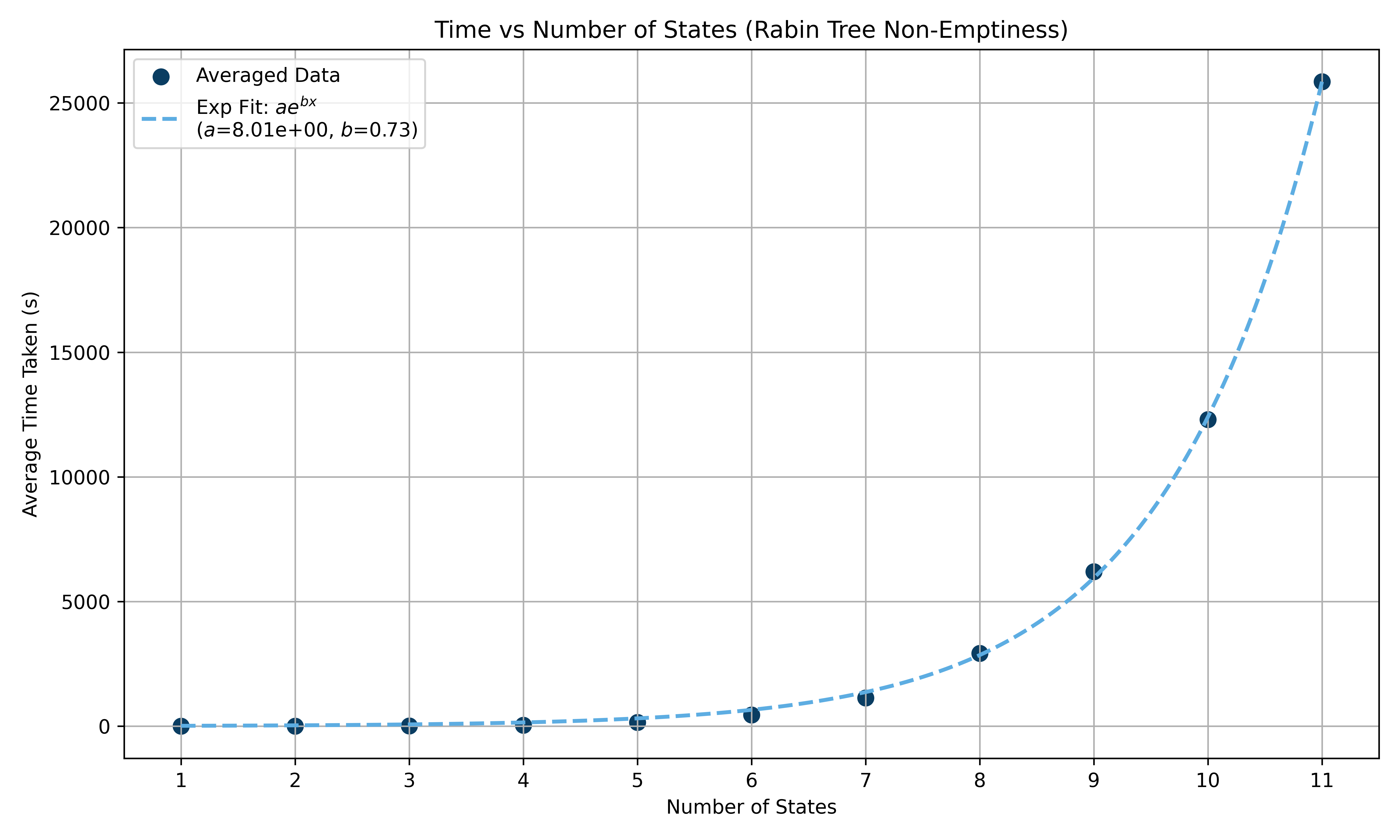}
    \caption{Runtime of the Rabin-based model checker as CGS size increases via the counter modulus \(m\).}
    \label{fig:rabin-parity-comp}
\end{figure}

\begin{figure}[H]
    \centering
    \includegraphics[width=0.7\linewidth]{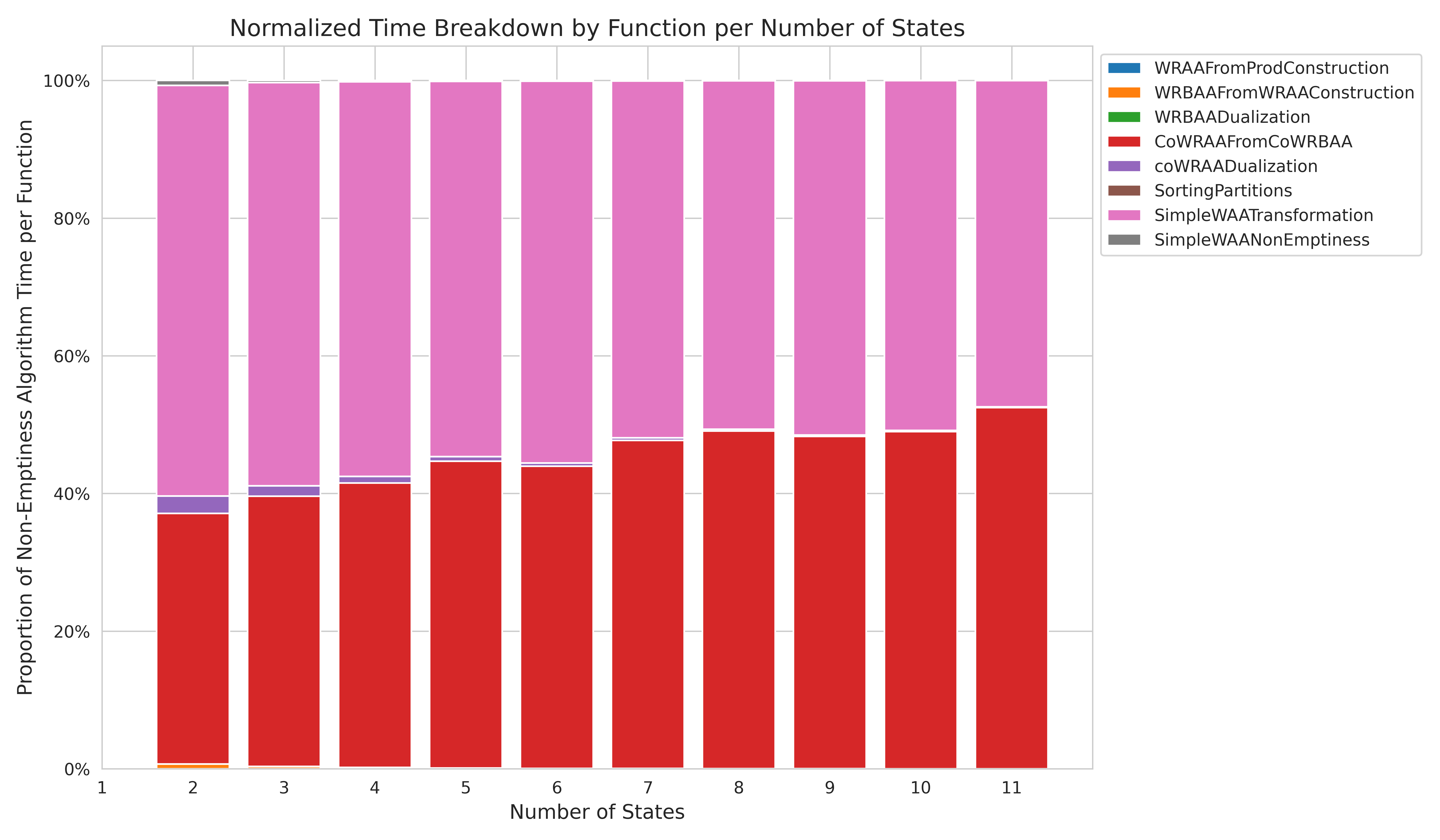}
    \caption{Normalised time spent per function of the Rabin non-emptiness algorithm.}
    \label{fig:time-by-function}
\end{figure}

Despite the hybrid formulation, evaluation revealed severe performance bottlenecks. As shown in Figure~\ref{fig:rabin-parity-comp}, the RTA approach becomes impractical beyond small model sizes (e.g. 7 hours for $m = 11$). Detailed profiling (Figure~\ref{fig:time-by-function}) shows the majority of runtime is spent on the automata transformations in the emptiness phase.
These results indicate that the Rabin-tree-based approach is not practical for real-world verification, even with the symbolic encodings used in the implementation, since the exponential blow-up in intermediate representations quickly renders the method infeasible. In contrast, the parity-game-based approach used in ATL\(^*\)AS avoids these bottlenecks and scales significantly bettwer, making it a more suitable approach for efficient ATL\(^*\) model checking in practice.

\end{document}